\documentclass[twocolumn,showpacs,prb,amsfonts,amsmath,amssymb,floatfix,groupedaddress]{revtex4-2}

\usepackage{color}
\usepackage{hhline}
\usepackage{mathrsfs}
\usepackage{graphicx}
\usepackage{dcolumn}
\usepackage{bm}
\usepackage{multirow}
\usepackage{booktabs}
\usepackage{afterpage}
\usepackage{amsmath}
\usepackage[inline]{asymptote}
\usepackage[caption=false]{subfig}
\usepackage[version=4]{mhchem}
\usepackage{multirow}

\everymath{\displaystyle}

\begin{document}

\title{Assessing the possible superconductivity in doped perovskite hydride \ce{KMgH3}: 
Effects of lattice anharmonicity and spin fluctuations}
\author{Shaocong Lu$^1$}
\thanks{shaocong.lu@phys.s.u-tokyo.ac.jp}
\author{Ryosuke Akashi$^2$}
\author{Mitsuaki Kawamura$^3$}
\author{Shinji Tsuneyuki$^1$}
\affiliation{$^1$Department of Physics, The University of Tokyo, Hongo, Bunkyo-ku, Tokyo 113-0033, Japan}
\affiliation{$^2$Quantum Materials and Application Research Center, 
              National Institutes for Quantum Science and Technology,
              2-10, Ookayama, Meguro-ku, Tokyo 152-0033, Japan}
\affiliation{$^3$Information Technology Center, The University of Tokyo, Tokyo, Tokyo 113-8658, Japan}

\date{\today}
\begin{abstract}
  The superconducting properties of 
uniformly hole-doped perovskite hydride \ce{KMgH3} 
with varying doping concentration and lattice parameter corresponding to 
different pressures were investigated from first principles. 
The superconducting transition temperature ($T_{\mathrm{c}}$) was predicted 
from the density functional theory for superconductors (SCDFT), where 
the effects of lattice anharmonicity and spin-fluctuation were considered and examined. 
Although lattice anharmonicity tends to suppress
superconductivity around the edge of dynamical stability, where the phase is 
stabilized due to anharmonic effects, $T_{\mathrm{c}}$ 
is enhanced. In the hole-doped \ce{KMgH3}, substantial 
spin-fluctuation (SF) effects were discovered, which  
counters the phonon-mediated pairing and decreases $T_{\mathrm{c}}$.
Such anomalously strong SF is evaluated for similar hydrides,
where the hydrogen 1-$s$ bands are isolated at the Fermi level, 
and its correlation with the electronics density of states was explored.
\end{abstract}

\maketitle

\section{Introduction}
Hydrogen-rich compounds, along with metallic hydrogen, have long been predicted to be promising candidates for 
realizing high-temperature superconductivity\cite{Ginzburg1969-ww, PhysRevLett.21.1748, PhysRevLett.92.187002}.
The discovery of superconductivity in sulfur hydride
under pressure in 2015\cite{Drozdov2015} again stimulated 
the enthusiasm for hydride superconductors. However, 
hydride superconductors which exhibit high transition temperatures 
($T_{\mathrm{c}}$) almost inevitably require a high pressure (e.g., 
$T_{\mathrm{c}}$ $\sim$ 203 K for \ce{H3S} under 200 GPa\cite{Drozdov2015}, 
$T_{\mathrm{c}}$ $\sim$ 260 (250) K for \ce{LaH10} under 200 (170) GPa\cite{PhysRevLett.122.027001, Drozdov2019}, 
$T_{\mathrm{c}}$ $\sim$ 224 (220) K for \ce{YH6} under 166 (183) GPa\cite{Troyan2021, Kong2021}). 
Despite being a challenging task, 
searching for hydride superconductors that 
retain a moderate value of $T_{\mathrm{c}}$ at a much lower or even 
near ambient pressure is still tempting. 
So far, some predictions on ambient pressure 
hydride superconductors have been proposed, such as 
$T_{\mathrm{c}}~\sim$ 54 K for metastable \ce{Al4H}, which adopts a 
perovskite-like crystal structure\cite{PhysRevB.108.054515}, and 
$T_{\mathrm{c}}$ between 45K to 80K for the family of \ce{Mg2XH6} compounds, where X = 
Rh, Ir, Pd or Pt\cite{Sanna2024}.

The theory of conventional superconductors was first given by 
Bardeen, Cooper, and Schrieffer\cite{PhysRev.108.1175} 
(BCS theory), according to which the superconducting transition temperature $T_{\mathrm{c}}$ can be estimated from
\begin{equation}
  T_{\mathrm{c}} = \Theta_{D}{\text{exp}}\left(-\frac{1}{N\left(\epsilon_\text{F}\right)V}\right),
\end{equation}
where $\Theta_D$, $N(\epsilon_\text{F})$, and $V$ denote the Debye temperature, 
electronic density of states (DOS) at the Fermi level, and 
the pairing interaction, respectively. Based on 
the BCS theory, materials with larger DOS at the Fermi surface
and stronger lattice vibrations are thought to possibly have 
a higher $T_{\mathrm{c}}$. So far, much effort has been put into the prediction 
of novel hydride superconductors, developing 
high-throughput screening or machine learning models
for the prediction relying on the database composed of materials satisfying
these two conditions\cite{Cerqueira2023, Choudhary2022, PhysRevB.104.054501, Wines2023}.

\begin{figure}[b]
  \centering
  \includegraphics[width=\columnwidth]{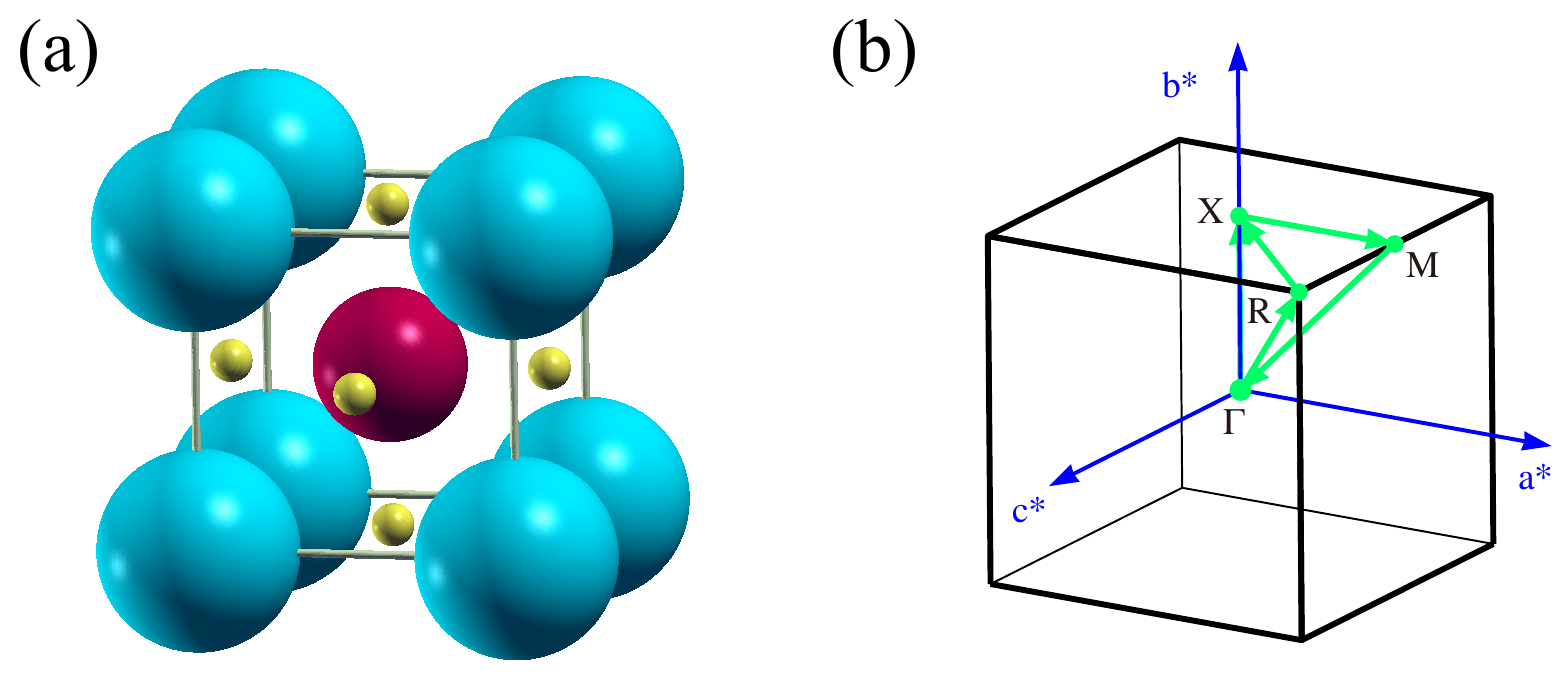}
  \caption{\label{structure_BZ}(a) Crystal structure of perovskite hydride \ce{KMgH3}. 
  The blue, red, and yellow balls represent K, Mg, H atoms respectively. 
  (b) Illustration of the Brillouin zone.}
\end{figure}

In this paper, we assess the possible superconductivity 
in hole-doped perovskite \ce{KMgH3} (Fig.~\ref{structure_BZ}). Studies on 
superconducting perovskite hydrides were few,
possibly due to the lack of stability in metallic perovskite hydrides.
Yagyu \textit{et al.} measured $T_{\mathrm{c}}$ of \ce{APdH3-x} (A = Sr, Ba), only to find out
they do not exceed 2K\cite{YAGYU2013109}. Cerqueira \textit{et al.}
predicted the $T_{\mathrm{c}}$ of \ce{KCdH3}, a 
metallic perovskite hydride at ambient pressure, 
to be 23.4K from the density functional theory for 
superconductors (SCDFT)\cite{https://doi.org/10.1002/adma.202307085}. 
Although \ce{KMgH3} is a known compound\cite{SCHUMACHER1990179},
the superconductivity possibility of it brought by doping
has not yet been explored. 
The parent compound \ce{KMgH3} is an insulator, while its valence states are wholly formed by the hydrogen 1\textit{s} orbitals. 
The hole doping is intriguing in that it should realize the metallic state 
strongly coupling to the hydrogen vibrations. 
Perovskite oxyhydrides \ce{ATiO2H} (A=K, Rb, Cs), 
which also share similar valence features, have been predicted to show the 
hole-doping induced superconductivity that coexists with nonzero polarization\cite{PhysRevMaterials.5.054802}. 
Such kinds of oxyhydrides were also reported to have been synthesized\cite{Goto2017, DenisRomero2014, Tassel2014, Sasahara2024}. 
The undoped \ce{KMgH3} already shows a excellent stability
from ambient pressure to 100GPa, and its highest optical phonon branches
can reach around 1,300 $\text{cm}^{-1}$ at ambient pressure.

Despite all the conditions favoring high $T_{\mathrm{c}}$, hydride 
superconductors may face some intrinsic drawbacks.
Hydrides usually feature strong lattice vibrations, 
giving rise to large lattice anharmonicity\cite{PhysRevB.44.7625, PhysRevB.29.6165},
which impact superconductivity\cite{PhysRevB.93.094525, PhysRevLett.111.177002, PhysRevLett.114.157004, PhysRevB.103.134305}.
Errea \textit{et al.} reported that electron-phonon coupling is reduced by 
approximately 30\% in \ce{H3S} under pressure\cite{PhysRevLett.114.157004}, while 
Sano \textit{et al.} showed that anharmonicity 
reduces the $T_{\mathrm{c}}$ of \ce{H3S} under 250 GPa by 12\%\cite{PhysRevB.93.094525}.
Errea \textit{et al.} also claimed that for palladium hydrides,
accounting for anharmonicity reduces the predicted $T_{\mathrm c}$ from 47 K 5 K, 
in better agreement with experimental values\cite{PhysRevLett.111.177002}.
As a matter of fact, lattice anharmonicity in the 
\ce{PdH} system has been extensively studied 
and is thought to be one of the reasons for the inverse isotope effect\cite{Ganguly1973, PhysRevLett.57.2955, PhysRevB.45.12405}.
Nevertheless, apart from its influence on $T_{\mathrm{c}}$, anharmonicity is also known
to possibly stabilize crystal structures. 
Errea \textit{et. al.} suggested 
that lattice anharmonicity is the key to the 
phase stability of $Fm\bar{3}m$ \ce{LaH10} under pressures between
129 GPa to 214 GPa\cite{Errea2020}, and $Im\bar{3}m$ \ce{H3S} at 155 GPa\cite{Errea2016}.

In addition, we raise the potential importance of spin fluctuation (SF) in the hydride metals. 
SF is considered to play against 
the phonon-mediated pairing between electrons, as the ferromagnetic correlation 
induced by the exchange coupling tends to destroy the Cooper pairs in the singlet state\cite{PhysRevLett.17.433}. 
In some transition metals with anomalously high magnetic susceptibility, 
$T_{\mathrm{c}}$'s predicted from 
a phonon-mediated pairing mechanism are known to yield 
overestimated values, which are assumed to be a 
consequence of neglecting the SF effects in these metals\cite{PhysRevLett.17.433, 10.1063/1.1709567, PhysRevLett.43.1256}.
Some of the authors of this paper have shown in their
previous study within the SCDFT framework that in elementary superconductors, SF 
reduces $T_{\mathrm{c}}$\cite{PhysRevB.101.134511, PhysRevB.102.214515}. 
The $T_\mathrm{c}$ reduction was substantial in 3\textit{d} systems like vanadium. 
According to the degree of the orbital radius, the hydride metals, formed by 
the extremely localized 1\textit{s} orbital, are likely to suffer from strong SF. 
Nevertheless, the SF-induced reduction of $T_{\mathrm{c}}$ in hydride superconductors 
has rarely been investigated, and, to our knowledge, there appears to be no paper 
on this matter, as the effects are considered trivial. 
In this paper, we will show a quantitative examination of such effects on 
hydride superconductors in which the conduction bands are mainly isolated 
hydrogen bands using the hole-doped \ce{KMgH3} as an example.

In this work, we present a comprehensive study on the stability and
superconductivity of hole-doped \ce{KMgH3} from first principles. 
By systematic examinations across different lattice parameters and
doping levels, and taking into account 
the effects of lattice anharmonicity and SF in the meantime, 
we hope to find the optimal conditions for $T_{\mathrm{c}}$ in this material , and to 
get a deeper understanding of hydride superconductors.

\section{Computational Methods}
The overall study was based on the Eliashberg theory\cite{eliashberg1960interactions} for the 
conventional phonon-mediated superconductors with the Migdal approximation\cite{migdal1958interaction}. 
$T_{\mathrm{c}}$ were estimated from McMillan-Allen-Dynes's formula (MAD)\cite{PhysRev.167.331, PhysRevB.12.905, DYNES1972615}:
\begin{eqnarray}
  \label{mcmillan}
  T_{\mathrm{c}} &= \frac{f_1f_2}{1.2}\text{exp}\left[-\frac{1.04\left(1+\lambda\right)}{\lambda-\mu^*-0.62\lambda\mu^*}\right], \\*
  \omega_{\text{log}}&=\textrm{exp}\left[\frac{2}{\lambda}\int_0^{\infty}\frac{d\omega}{\omega}\alpha^2F\left(\omega\right)\text{ln}\omega\right], \\*
\lambda &= 2\int_0^{\infty}\frac{d\omega}{\omega}\alpha^2F\left(\omega\right),
\end{eqnarray}
where $\omega_{\text{log}}$ is the logarithmic average of phonon frequencies, 
$\mu^*$ denotes the effective Coulomb repulsion, 
and $f_1$ and $f_2$ are the strong-coupling correction factor and
the shape correction factor, respectively, which are defined
in the following fashion:
\begin{eqnarray}
  f_1 &= \left[1+\left(\frac{\lambda}{\Lambda_1}\right)^{3/2}\right]^{1/3}, \\*
  f_2 &= 1 + \frac{\left(\bar{\omega}_2/\omega_{\text{log}}-1\right)\lambda^2}{\lambda^2+\Lambda_2^2},
\end{eqnarray}
with the two parameter $\Lambda_1$ and $\Lambda_2$ defined as:
\begin{eqnarray}
  \Lambda_1 &= 2.46\left(1+3.8\mu^*\right), \\*
  \Lambda_2 &= 1.82\left(1+6.3\mu^*\right)\left(\frac{\bar{\omega}_2}{\omega_{\text{log}}}\right),
\end{eqnarray}
and $\bar{\omega}_2$ denotes the 2nd order average 
of phonon frequencies:
\begin{eqnarray}
  \bar{\omega}_2 = \left[\frac{2}{\lambda}\int_0^{\infty}d\omega\alpha^2F\left(\omega\right)\omega\right]^{\frac{1}{2}}. 
\end{eqnarray}
$\alpha^2F(\omega)$ is the isotropic version of the
Eliashberg spectral function:
\begin{eqnarray}\label{a2Fomega}
  \alpha^2F(\omega) = &\frac{1}{N(\epsilon_\mathrm{F})}\int
  \frac{d\mathbf{k}d\mathbf{q}}{\Omega_{\text{BZ}}}
  \sum_{mn\nu}\lvert g_{mn\nu}\left(\mathbf{k},\mathbf{q}\right) \rvert
  \delta\left(\epsilon_{n\mathbf{k}}-\epsilon_{\mathrm{F}}\right) \notag\\*
  &\times\delta\left(\epsilon_{m\mathbf{k+q}}-\epsilon_{\mathrm{F}}\right)
  \delta\left(\hbar\omega_{\mathbf{q}\nu}-\hbar\omega\right),
\end{eqnarray}
where $\epsilon$ is the Kohn-Sham eigenvalues, $\omega_{\mathbf{q}\nu}$ denotes 
the phonon frequency with crystal momentum $\mathbf{q}$ in the branch $\nu$,
and $g_{mn\nu}(\mathbf{k},\mathbf{q})$ is the electron-phonon matrix element.
To avoid the artificial divergence occurring in the 
integral of $\alpha^2F(\omega)\mathbin{/}\omega$, we set the cutoff
in the small frequency region to be 0.0005 Ry whenever
such a term appeared in the integrand, and $\mu^*$ was taken to 
be 0.1.

We calculated the SF mediated electron-electron interaction 
within the adiabatic local density approximation\cite{PhysRevLett.76.1212, PhysRevB.90.214504, PhysRevB.102.214515}. 
In analogy to $\lambda$, the SF parameter, measuring the 
averaged SF strength over the Fermi surface, is defined as
\begin{eqnarray}
  \label{mu_s_def}
  \mu_s = \frac{1}{N(\epsilon_F)}\sum_{\mathbf{k}\mathbf{k}^{'}nn^{'}}
  V_{n\mathbf{k}n^{'}\mathbf{k}^{'}}^{\mathrm{SF}}
  \delta(\epsilon_{n\mathbf{k}})\delta(\epsilon_{n^{'}\mathbf{k}^{'}}),
\end{eqnarray}
where $V_{n\mathbf{k}n^{'}\mathbf{k}^{'}}^{\mathrm{SF}}$ is the 
effective interaction of SF, which is calculated in the following way:
\begin{eqnarray}
  V^{\mathrm{SF}}_{n\mathbf{k}n^{'}\mathbf{k}^{'}} = &\frac{2}{\pi}\int d \omega \frac
  {|\epsilon_{n\mathbf{k}}|+|\epsilon_{n^{'}\mathbf{k^{'}}}|}
  {\left( |\epsilon_{n\mathbf{k}}|+|\epsilon_{n^{'}\mathbf{k^{'}}}| \right)^2+\omega^2} \notag \\*
  & \times \Lambda^{\mathrm{SF}}_{n\mathbf{k}n^{'}\mathbf{k^{'}}}(i\omega).
\end{eqnarray}
$\Lambda^{\mathrm{SF}}_{n\mathbf{k}n^{'}\mathbf{k^{'}}}$ contains the effective interaction of SF, and can be derived from: 
\begin{eqnarray}
  \Lambda^{\mathrm{SF}}(\mathbf{r},\mathbf{r}',i\omega) = 
  &- \int \int d\mathbf{r}_1 d\mathbf{r}_2 I^{\mathrm{XC}}(\mathbf{r},\mathbf{r}_{\mathrm{1}}) \notag \\*
  &\times \Pi(\mathbf{r}_{1},\mathbf{r}_{2},i\omega)I^{\mathrm{XC}}(\mathbf{r}_{2},\mathbf{r}'), 
\end{eqnarray}
where $\Pi$ is the spin susceptibility and $I^{\mathrm{XC}}$ denotes the second-order 
functional derivative of the exchange-correlation energy with respect to the spin density. 
See Refs.~\cite{PhysRevB.101.134511, PhysRevB.102.214515, PhysRevB.90.214504} for more information.

The Kohn-Sham (KS) self-consistent calculations for charge densities were performed 
on a $12 \times 12 \times 12$ \textbf{k} grid. The uniform hole doping 
was realized by subtracting electrons from unit cells, and adding a 
compensating jellium background. 
Calculations of phonon energies $\omega_{\mathbf{q}\nu}$ and phonon line widths
$\gamma_{\mathbf{q}\nu}$ and the electron-phonon matrix element 
$g_{mn\nu}(\mathbf{k},\mathbf{q})$ within the harmonic 
approximation (HA) were performed using the density functional perturbation theory (DFPT) scheme 
implemented in QUANTUM ESPRESSO (QE)\cite{Giannozzi_2009, Giannozzi_2017} 
on a $6\times6\times6$ \textbf{q} point mesh. 
The \textbf{q} grid was half-shifted in order to avoid the divergence regarding the 
double delta integral in Eq.~(\ref{a2Fomega}). 
We used the ultra-soft pseudopotentials\cite{PhysRevB.41.7892} 
with PBE exchange and correlation functionals\cite{PhysRevLett.77.3865}.
The kinetic energy cutoff for wavefunctions and charge densities were
set to 100 Ry and 800 Ry, respectively. 
The optimized tetrahedron method\cite{PhysRevB.89.094515} was used to avoid 
broadening dependence. In the calculations of electron-phonon coupling (EPC), 
the derivative of the self-consistent potential was evaluated on the same \textbf{q} grid,
whereas the KS energies $\epsilon_{n\mathbf{k}}$ were computed in a non-self-consistent calculation 
with a $24\times24\times24$ $k$ grid. 

One approach to dealing with anharmonicity is using 
the self-consistent phonon method (SCPH)\cite{PhysRevB.92.054301}. 
In this approach, one renormalizes the harmonic phonon frequencies 
through the phonon self-energy. 
We used ALAMODE\cite{Tadano_2014} to do the SCPH calculations.
In the SCPH calculations, we first prepared a $2\times2\times2$ 
supercell containing 40 atoms and generated more than 160 configurations for each system by adding 
random displacements between 0.05 to 0.2 \AA, and calculated corresponding inter-atomic
forces using QE with the same pseudopotentials as mentioned before and a $8\times8\times8$ k-point mesh. 
The inter-atomic force constants (IFCs) were then fitted by the least squares method to
minimize the sum of residual squares. The IFCs were fitted 
up to the sixth order for a robust estimation of IFCs. 
For the second-order IFCs, the cutoff range included the whole 
supercell, while for the rest the cutoff range were set to 
include only the nearest neighbors.
The average fitting error was 0.925\%. The SCPH equations were solved on
a $4\times4\times4$ mesh at 0 K, and we extracted the anharmonic corrections
to the second-order IFCs. With the IFCs containing anharmonic corrections, 
we obtained the renormalized phonon frequencies, along with 
the renormalized Eliashberg spectral function $\alpha^2F(\omega)$, 
from which we calculated $\lambda$ and $\omega_{\text{log}}$.
The dynamical stability is judged from the phonon spectral.

Superconducting $T_\mathrm{c}$ was finally calculated without empirical parameters 
on the basis of SCDFT\cite{PhysRevB.72.024545} 
using the superconducting toolkit\cite{PhysRevB.101.134511}.
The random phase approximation (RPA)\cite{PhysRev.106.364} 
was applied in the computation of the screened Coulomb interaction, 
where the frequency dependence was retained for including the 
dynamical screening via the plasmon oscillation\cite{Takada1978, PhysRevLett.111.057006}.
The KS orbitals were calculated on a $6\times6\times6$ $k$ mesh. 
When calculating the superconducting gap equation, 
we included 50 energy bands, and adopted the for phononic part 
the exchange-correlation kernel SPG2020\cite{PhysRevLett.125.057001}.

\section{Results and Discussion}
\subsection{Electronic structure}
\begin{figure}[!h]
  \centering
  \includegraphics[width=\columnwidth]{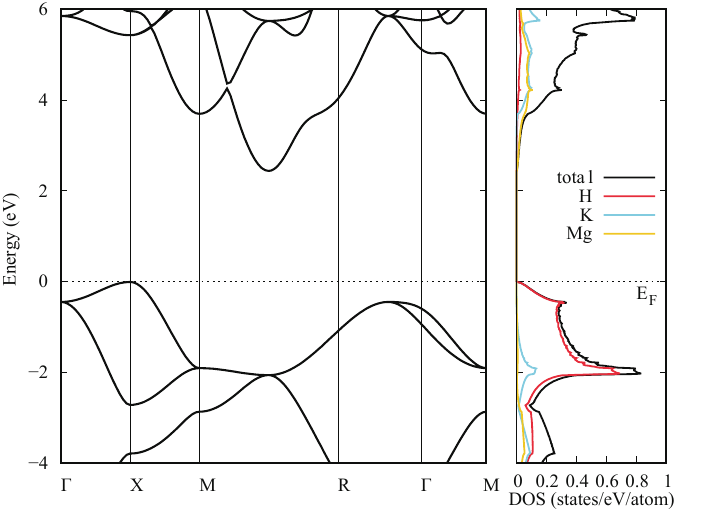}
  \caption{\label{e_structure}Electronic bands structure and DOS of \ce{KMgH3} beforedoping at ambient pressure.}
\end{figure}

In Fig.~\ref{e_structure} we plotted the electronic band structure 
and projected DOS of \ce{KMgH3}, without doping, at ambient pressure. 
The band gap is approximately 2.4 eV, whereas the valence states are dominated by 
hydrogen 1\textit{s} orbitals. 
The valence states of \ce{KMgH3} show a peak in the electronic DOS, 
whose value reaches about 
3.8 $\text{states}\cdot\text{eV}^{-1}\cdot u.c.^{-1}$. For comparison, 
the DOS at the Fermi level of $Im\bar{3}m$ \ce{H3S} under 250GPa is 
about 0.7 $\text{states}\cdot\text{eV}^{-1}\cdot u.c.^{-1}$\cite{PhysRevB.93.094525}. 
Below the Fermi level, the DOS monotonically increases, and reaches 
a maximum at around -2 eV relative to the Fermi level, which makes it an ideal choice for hole-doping. 

\subsection{Stability}
\begin{figure}[!t]
  \centering
  \includegraphics[width=\columnwidth]{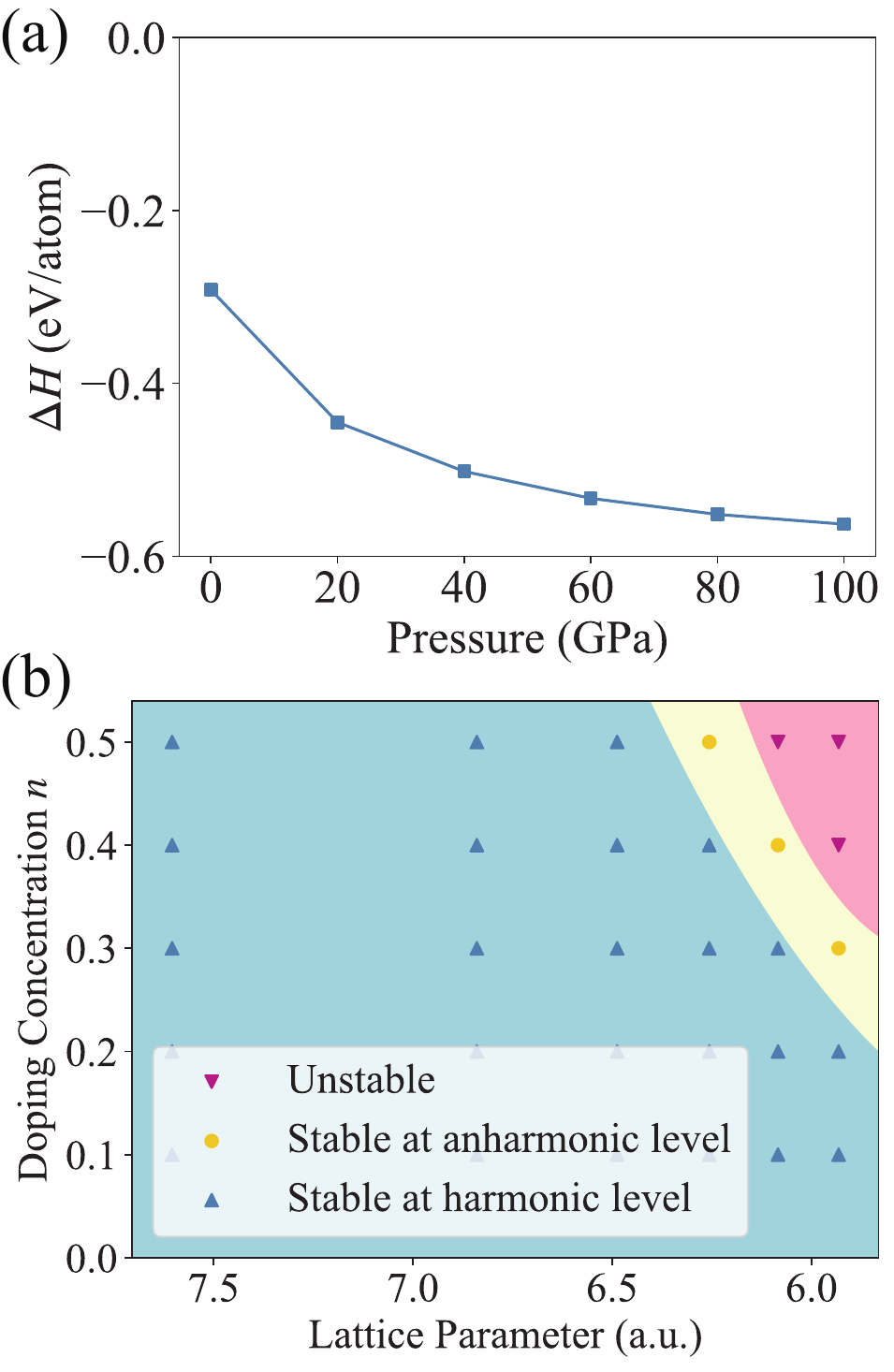}
  \caption{\label{enthalpy}(a) Calculated enthalpy difference (relative to \ce{KH} in the $Fm\bar{3}m$ phase 
  and \ce{MgH2} in the $P4_2/mnm$ phase) of \ce{KMgH3} 
  from ambient pressure to 100 GPa. 
  (b) The dynamical stability of the cubic phase of hole doped \ce{KMgH3} 
  with respect to doping concentration and lattice parameter. 
  Smaller lattice parameters indicates higher pressures.
  A lattice parameter of 7.6 (5.8) a.u. are equal to the lattice parameter 
  of the undoped compound under ambient (100GPa) pressure.
  }
\end{figure}

The thermodynamical stability of the pristine \ce{KMgH3} was confirmed by calculations 
of enthalpies relative to the binary hydrides \ce{KH} and \ce{MgH2} presented in Fig.~\ref{enthalpy}(a). 
We hence assume that it can stay thermodynamically stable with small doping, especially at higher pressures. 
The dynamical stability of the cubic phase
of doped \ce{KMgH3} was examined by first calculating
phonon frequencies within the harmonic approximation
with DFPT, and checking if imaginary phonon modes appeared.
These harmonic phonon frequencies are further renormalized
through SCPH iterations, and the ones still showing imaginary
modes are deemed to be unstable. 

\begin{figure*}[!ht]
  \centering
  \includegraphics[width=2\columnwidth]{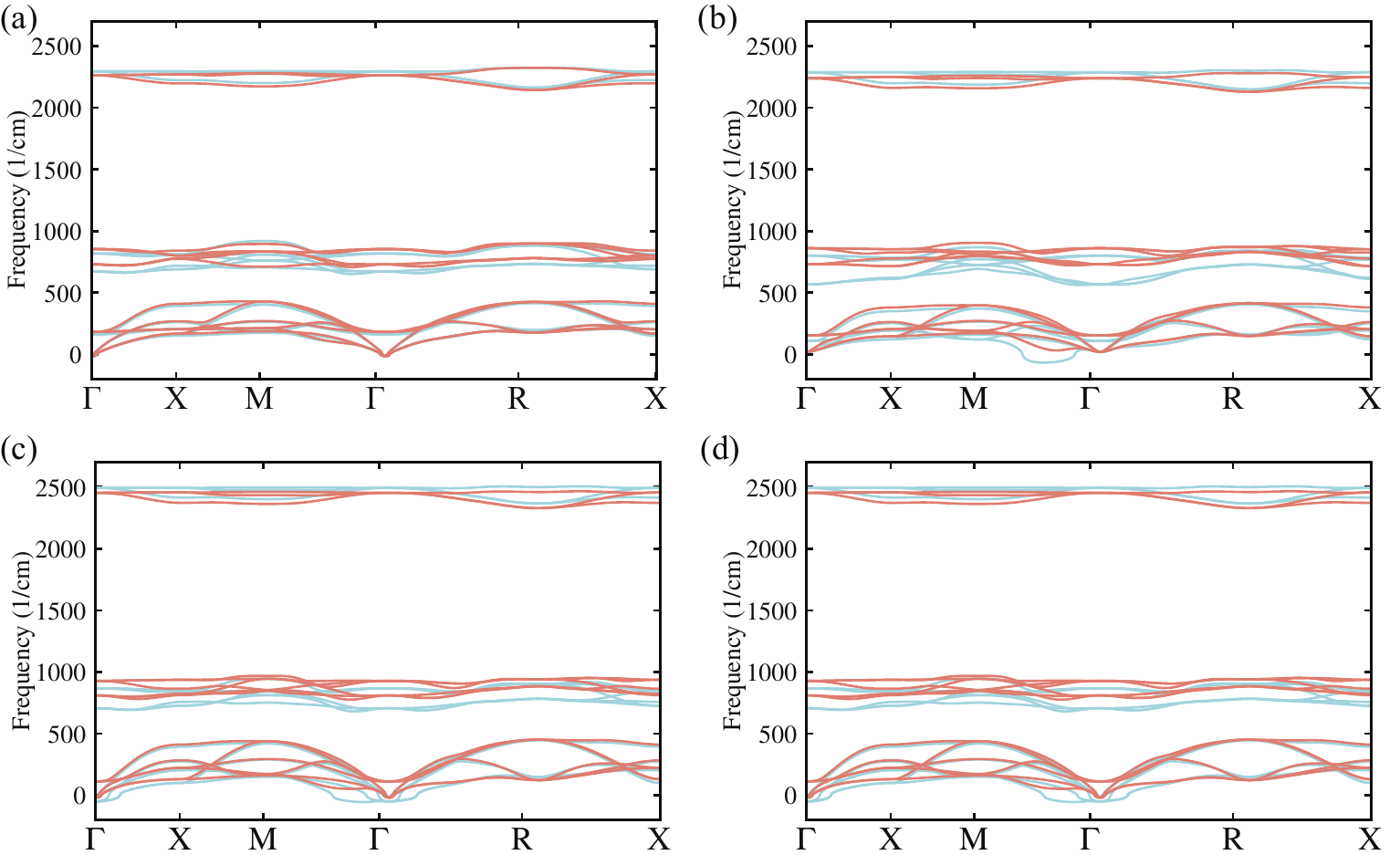}
  \caption{\label{dispersion}Phonon dispersions of uniformly hole-doped \ce{KMgH3} with lattice constant and 
  doping concentration respectively equal to (a) 6.26 a.u. and 0.4. (b) 6.26 a.u. and 0.5. 
  (c) 6.08 a.u. and 0.3. (d) 6.08 a.u. and 0.4. The blue lines are the harmonic frequencies, 
  and the red lines are the anharmonic phonon frequencies from SCPH.}
\end{figure*}

The phase diagram of doped \ce{KMgH3} is shown in Fig.~\ref{enthalpy}(b), 
where doping concentration \textit{n} denotes the 
number of electrons substracted per unit cell. 
The dynamical stability gradually decreases with
increasing pressure and doping concentration. Within
a narrow range, lattice anharmonicity helps to restore
the phase stability.
Fig.~\ref{dispersion} shows some typical phonon dispersions
with both harmonic and anharmonic frequencies.
The imaginary modes which appear around $\Gamma$ point are
hardened by anharmonicity, which indicates the dynamical 
stability brought about by anharmonic effects. Apart from this, 
the middle branches are mostly affected by anharmonicity, with
a universal enhancement in frequency of 5.71\% on average, while the frequency 
shift among the highest branches are on average only +0.43\%.

The impact of pressure is most significant on the highest branches.
Meanwhile, the other branches are also hardened by the increase
of pressure, along with the appearance of imaginary modes in the acoustic branches.
When the lattice parameter shrinks from 7.6 a.u. to 5.8 a.u., 
the average phonon frequencies of the three highest optical 
branches are strengthened from 1200 $\mathrm{cm^{-1}}$ to 2700 $\mathrm{cm^{-1}}$.
The middle branches are hardened from 600 $\mathrm{cm^{-1}}$ to about 1000 $\mathrm{cm^{-1}}$ on average.
On the other hand, the impact of doping is most obvious in 
the middle branches, leaving the highest branches almost untouched.
Similar to pressure, increasing \textit{n} also tends to weaken the dynamical stability.

\subsection{Superconductivity}
\begin{figure}
  \centering
  \includegraphics[width=\columnwidth]{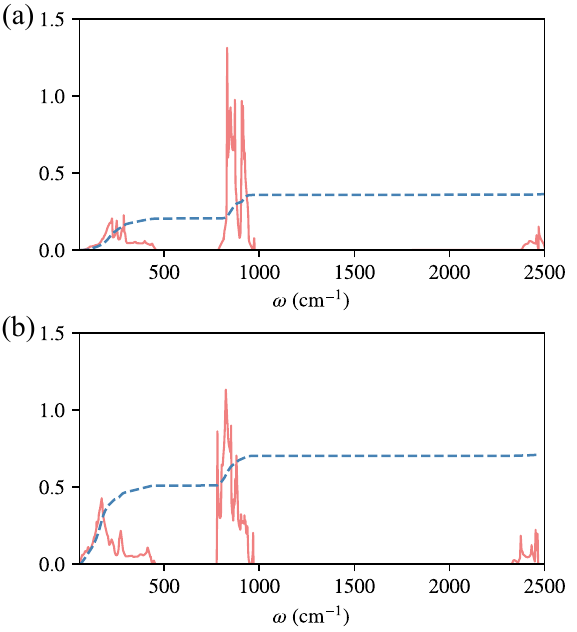}
  \caption{\label{a2f}The Eliashberg spectral functions when lattice constant is 6.08 a.u., and 
  doping concentration is (a) 0.3. (b) 0.4. The red solid lines are the spectal functions, and
  the blue dashed lines are integrated value of $\lambda$.}
\end{figure}

\begin{figure*}[!t]
  \centering
  \includegraphics[width=2\columnwidth]{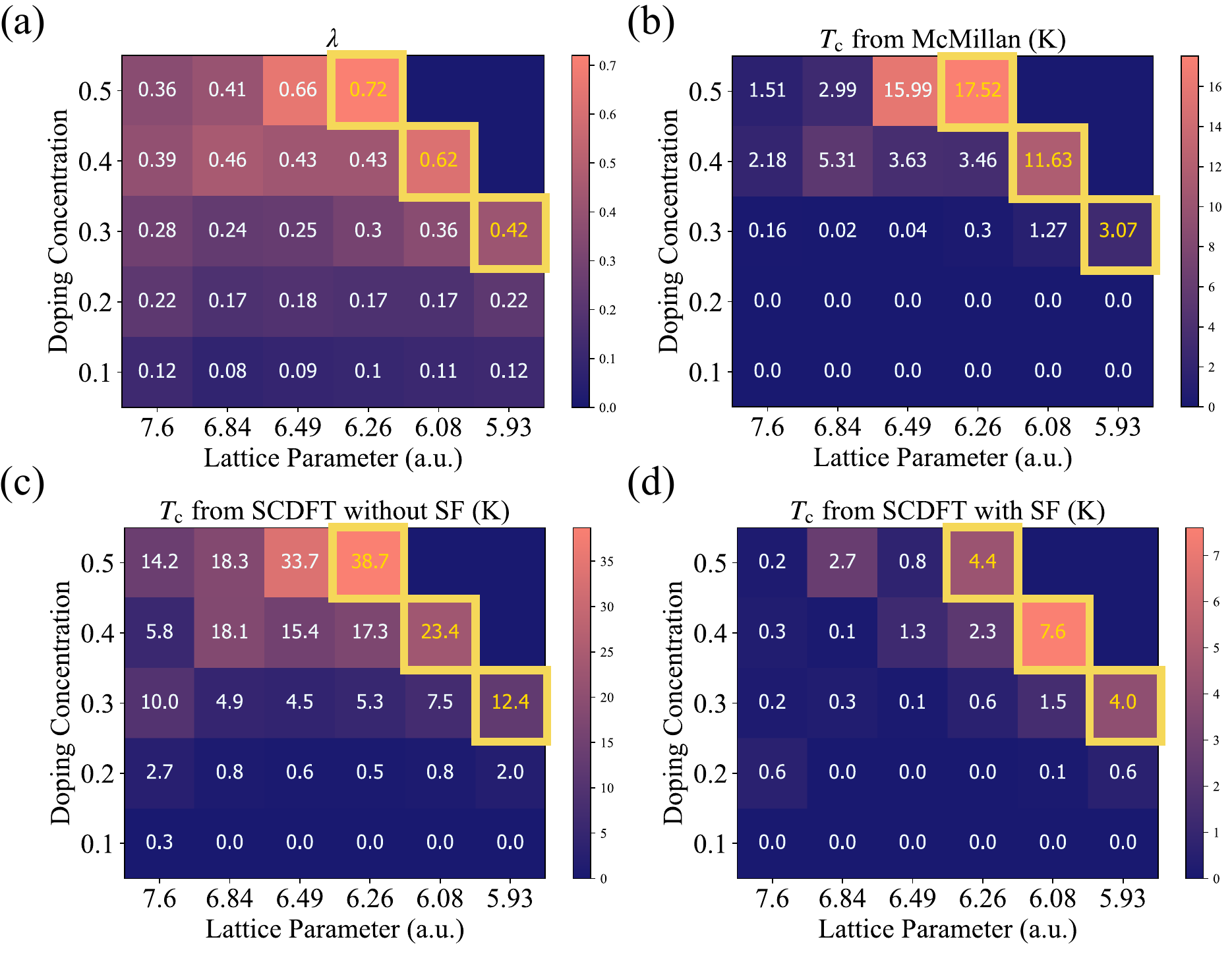}
  \caption{\label{lambda_tc}Lattice constant and doping concentration dependence of 
  (a) EPC parameter $\lambda$. 
  (b) $T_{\mathrm{c}}$ estimated from MAD formula with $\mu^*=0.1$. 
  (c) $T_{\mathrm{c}}$ predicted from SCDFT without SF. 
  (d) $T_{\mathrm{c}}$ predicted from SCDFT with SF included. 
  The highlighted parts indicate the ones that are only stable 
  at anharmonic level. }
\end{figure*}

For the stable systems, we performed calculations
of electron-phonon coupling with the renormalized 
phonon frequencies. 
Fig.~\ref{a2f} contains two typical plots of Eliashberg functions. 
The coupling is strongest among middle and lower frequencies, while in the
high-frequency branches the coupling is negligible, indicating
that such vibrations of hydrogen do not participate
in EPC. Consequently, the pressure effects on $\lambda$ appeared to be
ambiguous in the lower doping case, since 
pressure mostly affects the highest branches of phonon modes, 
which are barely coupled with electron, and have almost no 
influence on $\lambda$. From Fig.~\ref{a2f} we see the contributions
to $\lambda$ almost entirely come from the middle and low branches of phonon modes. \\
We calculated the EPC strength $\lambda$ and plotted them in 
Fig.~\ref{lambda_tc}(a). $\lambda$ generally increases with 
the doping concentration, because 
the Fermi surface in this compound is brought about by 
doping entirely, and with more doping we can expect a 
larger coupling strength due to the increased $N(\epsilon_F)$. 
The EPC is insignificant when $n\le0.2$, only giving $\lambda$'s 
mostly smaller than 0.2. Therefore, we do not expect 
superconductivity from this small doping region. 
The strongest coupling happens when $a=6.26$ a.u. 
and $n=0.5$, where $\lambda$ reaches 0.72.
Overall, the hole-doped \ce{KMgH3} is far from a strong coupling 
material as we have expected. Furthermore, 
by including lattice anharmonicity, $\lambda$ is slightly 
decreased. Such suppression of EPC in hydrides 
due to anharmonicity is in qualitative agreement with 
previous researches\cite{PhysRevB.93.094525, PhysRevB.103.134305}.
This is because the hardening moves the coupling phonon frequencies farther 
from the optimal regime of $\mathbin{\sim}7T_\mathrm{c}$\cite{Bergmann1973, PhysRevB.96.100502, PhysRevB.103.134517}.

Transition temperatures were estimated from Eq.~(\ref{mcmillan}) for the stable 
systems whose $\lambda$ is no less than 0.1, and the results are
plotted in Fig.~\ref{lambda_tc}(b). We see that there is almost no superconductivity
found for $n\le0.3$ except at high pressure ($a=5.8$ a.u. and $n=0.3$), 
as a result of minor $N(\epsilon_{\textrm{F}})$ values. 
Just like $\lambda$, $T_\textrm{c}$ benefits from the larger $N(\epsilon_{\textrm{F}})$
rise by the increasing doping concentration. 
The highest $T_\textrm{c}$ (K) occurs under the conditions where 
$a=6.26~a.u.$ and $n=0.5$, which would have been considered unstable
if only treated within HA. 

\begin{figure}[!t]
  \centering
  \includegraphics[width=\columnwidth]{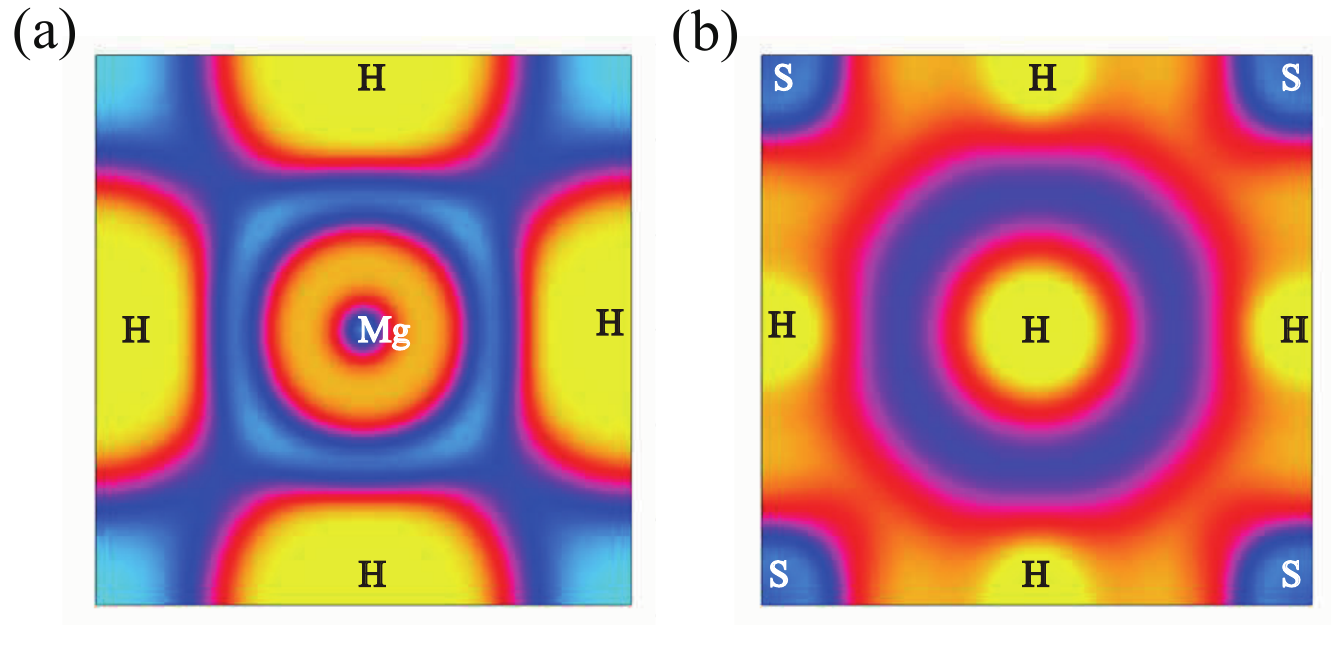}
  \caption{\label{elf}Electron localization function of (a) \ce{KMgH3}, where lattice constant is 
  6.26 a.u. and doping concentration is 0.5, in the Mg-H plane. 
  (b) $Im\bar{3}m$ \ce{H3S}. The cyan, magenta, and yellow 
  colours represent high, medium, and low density, respectively. 
The two figures are made with VESTA\cite{Momma:db5098}.}
\end{figure}

In order to get rid of the arbitrariness brought about by 
the empirical choice of $\mu^*$, we also performed SCDFT calculations 
for all the systems. Fig.~\ref{lambda_tc}(c) and Fig.~\ref{lambda_tc}(d) compare 
transition temperatures predicted from SCDFT without and with SF. 
Without SF, the general trends in $T_{\mathrm{c}}$ qualitatively agree with 
those in the $T_{\mathrm{c}}$ from MAD, because in both cases only phonon and 
Coulomb contributions were treated. 
The inclusion of SF caused a substantial decrease in $T_{\mathrm{c}}$. 
The highest transition temperature no longer happens when the lattice parameter is 6.26 $a.u.$ 
and $n$ is 0.5; for that point $T_{\mathrm{c}}$ drops from 38.7 K to only 4.4 K. 
Overall, the highest $T_{\mathrm{c}}$ becomes 7.6 K.
The reduction of $T_{\mathrm{c}}$ indicates a strong SF in the doped \ce{KMgH3}, 
which could result from the localized electrons orbitals\cite{PhysRevB.102.214515}. 
In Fig.~\ref{elf}(a), 
we plotted the electron localization functions within 
the Mg-H plane for doped \ce{KMgH3}. 
Despite the doping and the applied pressure, the electronic orbitals are still largely 
localized, contributing to a considerable part of the magnetic exchange-correlation 
kernel, which results in a lower $T_{\mathrm{c}}$. In comparison, a plot of ELF for 
$Im\bar{3}m$ \ce{H3S} in the (1, 0, 0) plane is shown in Fig.~\ref{elf}(b). The electrons 
are well delocalized between S atoms and their nearest H atoms, which may 
explain the minor reduction on $T_{\mathrm{c}}$ from 203 K to 190 K 
due to SF in \ce{H3S}\cite{PhysRevMaterials.6.114802}.
By applying pressure or increasing doping concentration, it is possible 
to enhance $T_{\mathrm{c}}$ through increasing the EPC strength and lowering the SF effect, 
because of the more overlapped orbitals. However, maintaining the phase stability 
under such a situation could be a difficult problem.\\
Interestingly, compared with neighboring systems stable within HA, 
$T_{\mathrm c}$ was enhanced in all systems stabilized by anharmonic effects 
where the possibility of superconductivity is usually not considered. 
The present calculations suggest the possibility of discoviering new 
superconductors in such systems stabilized by the anharmonic effefct.

\subsection{Spin fluctuation}
In this subsection we dwell more into the possible origin 
of such strong SF effects in the doped \ce{KMgH3}. 

\begin{figure}[!t]
  \centering
  \includegraphics[width=\columnwidth]{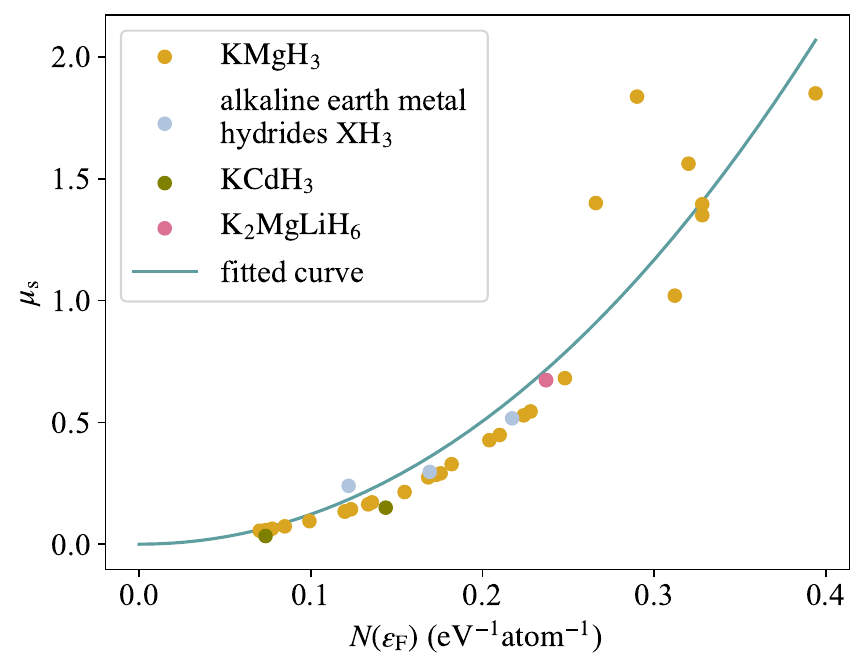}
  \caption{\label{hpdos_mu}SF strength $\mu_s$ with respect to the DOS at the Fermi level $N(\epsilon_{\mathrm{F}})$. 
  The solid circles are all the stable doped \ce{KMgH3}, alkaline earth metal hydrides \ce{XH3}, X being Mg, Ca and Sr, 
  perovskite hydride \ce{KCdH3} before and after doping, and Li-doped \ce{KMgH3}, respectively. 
  The solid line is the fitted curve according to Eq.~(\ref{approxmu}).}
\end{figure}

In Fig.~\ref{hpdos_mu}, we plotted the correlation between 
the SF strength $\mu_s$, defined in Eq.~\ref{mu_s_def}, and $N(\epsilon_{\mathrm{F}})$. 
In this figure, we include all the stable phases 
of doped \ce{KMgH3}. To confirm the generality of the trend, we also calculated $\mu_\mathrm{s}$ 
of other hydrides having valence states formed almost exclusively by 
H-1\textit{s} orbitals, either stable or unstable\cite{supp}. 
This feature is rarely seen in experimentally confirmed 
or theoretically predicted hydride superconductors, since the H-1$s$ bands usually only make a small 
contribution to the total $N(\epsilon_{\mathrm{F}})$. 
If we assume that the spin-spin interaction is constant, 
the effective interaction averaged over the Fermi surface can 
be estimated as:
\begin{eqnarray}
  \label{approxV}
  V_{n\mathbf{k}n^{'}\mathbf{k}^{'}}^{\mathrm{SF}}
  \approx \frac{s_1N(\epsilon_{\mathrm{F}})}{1-s_2N(\epsilon_{\mathrm{F}})},
\end{eqnarray}
where $s_1$ and $s_2$ are constants among the same type of hydrides. Combining 
Eq.~(\ref{approxV}) and~(\ref{mu_s_def}), we have the following approximation
\begin{eqnarray}
  \label{approxmu}
  \mu_s \approx \frac{s_1N(\epsilon_{\mathrm{F}})^2}{1-s_2N(\epsilon_{\mathrm{F}})}.
\end{eqnarray}
In materials like doped 
\ce{KMgH3}, $\mu_s$ grows substantially when the $N(\epsilon_{\mathrm{F}})$ becomes larger, and 
the correlation can still be well fitted to Eq.~(\ref{approxmu}), 
in spite of the crude approximation. 
The largest $\mu_s$ can reach as high as 1.85. Since doped \ce{KMgH3} 
is only a medium to low coupling superconductor, such $\mu_s$ is 
expected to largely reduce $T_{\mathrm{c}}$, or suppress superconductivity completely 
even at low $N(\epsilon_{\mathrm{F}})$, because $\lambda$ is also approximately proportional to $N(\epsilon_{\mathrm{F}})$. 
When $\mu_s$ is close or even larger than $\lambda$, 
$T_{\mathrm{c}}$ could be greatly or even completely 
suppressed (e.g., in elemental scandium)\cite{PhysRevB.101.134511}, 
which is what happens in the case of doped \ce{KMgH3}.

This phenomenon indicates a possible dilemma for hydride superconductors. While one 
attempts to make use of the H-1$s$ orbitals in 
hydride superconductors to seek a strong EPC, such hydrides will 
possibly suffer from the suppression of $T_{\mathrm{c}}$ resulted from large SF effects.

\section{Summary}
We have examined the superconducting properties, including 
the EPC strength, SF effects,  and superconducting transition temperature, 
of uniformly hole-doped perovskite hydride \ce{KMgH3} under 
varying doping concentration and lattice parameters, with lattice 
anharmonicity included. 
The highest $T_{\mathrm{c}}$ predicted by SCDFT is 7.6 K. 
The suppressed $T_{\mathrm{c}}$ indicates that the doped \ce{KMgH3} 
could be an interesting material where strong SF effects are expected, which reduce $T_{\mathrm{c}}$. 
The exceptional suppression of superconductivity caused by SF could also be expected 
to happen in other hydride superconductors, where the hydrogen 1-$s$ bands are 
isolated at the Fermi level, which calls for additional attention 
in the current search for hydride superconductors.
We also found that $T_\textrm{c}$ is enhanced when the system 
is on the edge of stability and is stabilized by 
lattice anharmonicity. Along with the suppression effects 
of anharmonicity on $T_\textrm{c}$, our finding again emphasizes 
the important role of lattice anharmonicity in the study of 
hydride superconductors. 

\begin{acknowledgments}
We would like to express our gratitude to Dr. Terumasa Tadano, 
Dr. Takahiro Ishikawa and Dr. Tomohito Amano for enlightening disccusions and valuable suggestions. 
The calculations in this research were carried out on the supercomputer Ohtaka
of Institude of Solid State Physics, The University of Tokyo.
This work was supported by JSPS KAKENHI Grant No. 23K03313 from 
Japan Society for the Promotion of Science (JSPS).
\end{acknowledgments}

\bibliography{ref.bib}

\begin{thebibliography}{70}
\bibitem [{sup()}]{supp}%
  \BibitemOpen
  \href@noop {} {}\bibinfo {note} {See {S}upplemental {M}aterial at
  {URL}-will-be-inserted-by-publisher for more detailed information, which also
  includes Refs. \cite{ KRESSE199615, PhysRevB.47.558, PhysRevB.59.1758,
  PhysRevB.54.11169, VANSETTEN201839, PhysRevB.88.085117}.}\BibitemShut {Stop}%
\end{thebibliography}

\onecolumngrid
\newcounter{figureSM}
\newcounter{tableSM}
\setcounter{figure}{0}
\setcounter{table}{0}
\renewcommand{\thefigure}{\textsc{S}-\arabic{figure}}
\renewcommand{\thetable}{\textsc{S}-\arabic{table}}
\newpage
\makeatletter
\begin{center}
  {\large{\bf Supplementary Materials}}
\end{center}
\setcounter{page}{1}
\subsection{Doped $\mathbf{KMgH_3}$}
\begin{figure}[!ht]
  \centering
  \includegraphics[width=\columnwidth]{"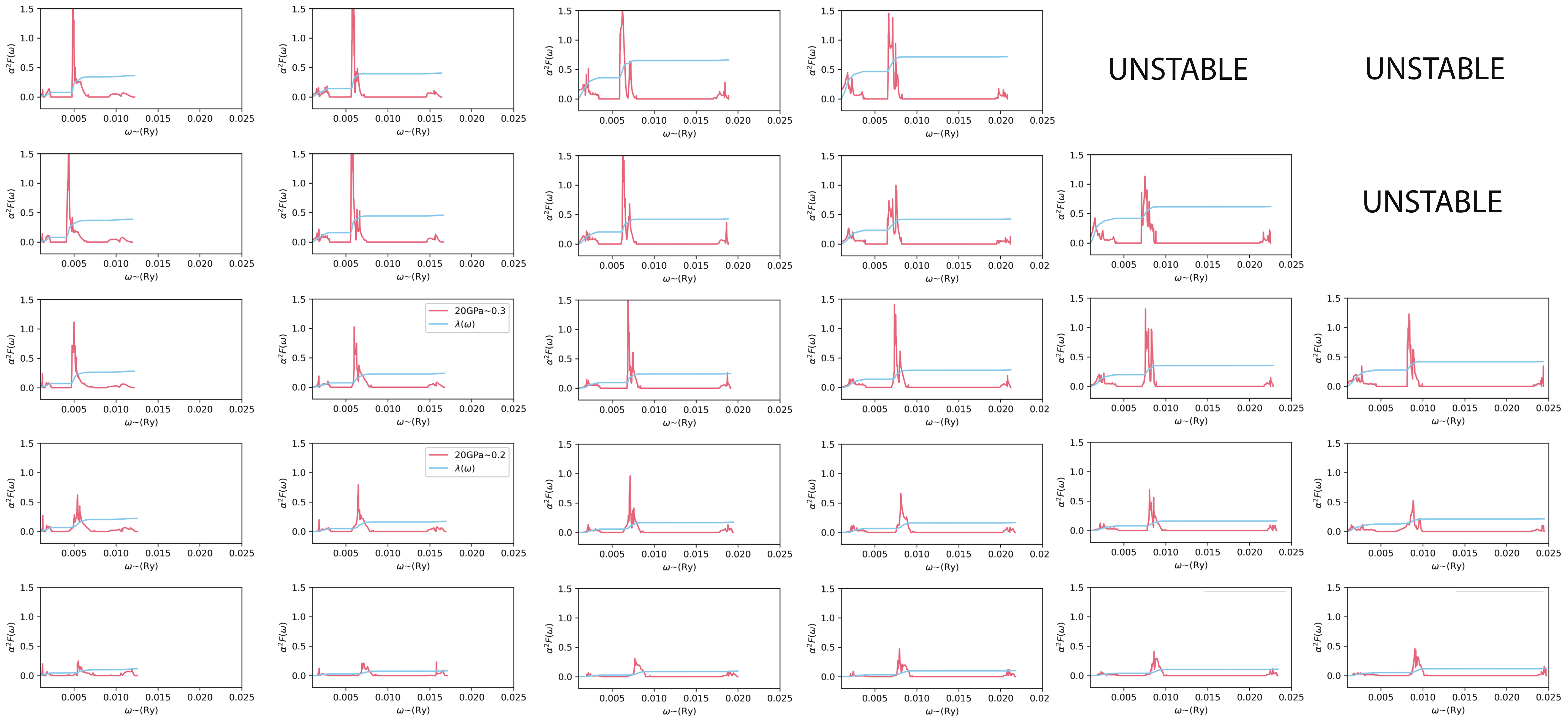"}
  \caption{The Eliashberg spectral functions $\alpha^2F(\omega)$ and the electron-phonon 
  coupling parameter $\lambda$. 
  The red lines indicate the $\alpha^2F(\omega)$ , and the blue lines are integrated $\lambda$. 
  Columns from left to right correspond to 
  lattice parameters of 7.60, 6.84, 6.49, 6.26, 6.08 and 5.93 bohr. Rows from bottom to 
  top correspond to doping concentrations of 0.1, 0.2, 0.3, 0.4 and 0.5.}
  \label{a2F}
\end{figure}
\begin{figure}[!ht]
  \centering
  \includegraphics[width=\columnwidth]{"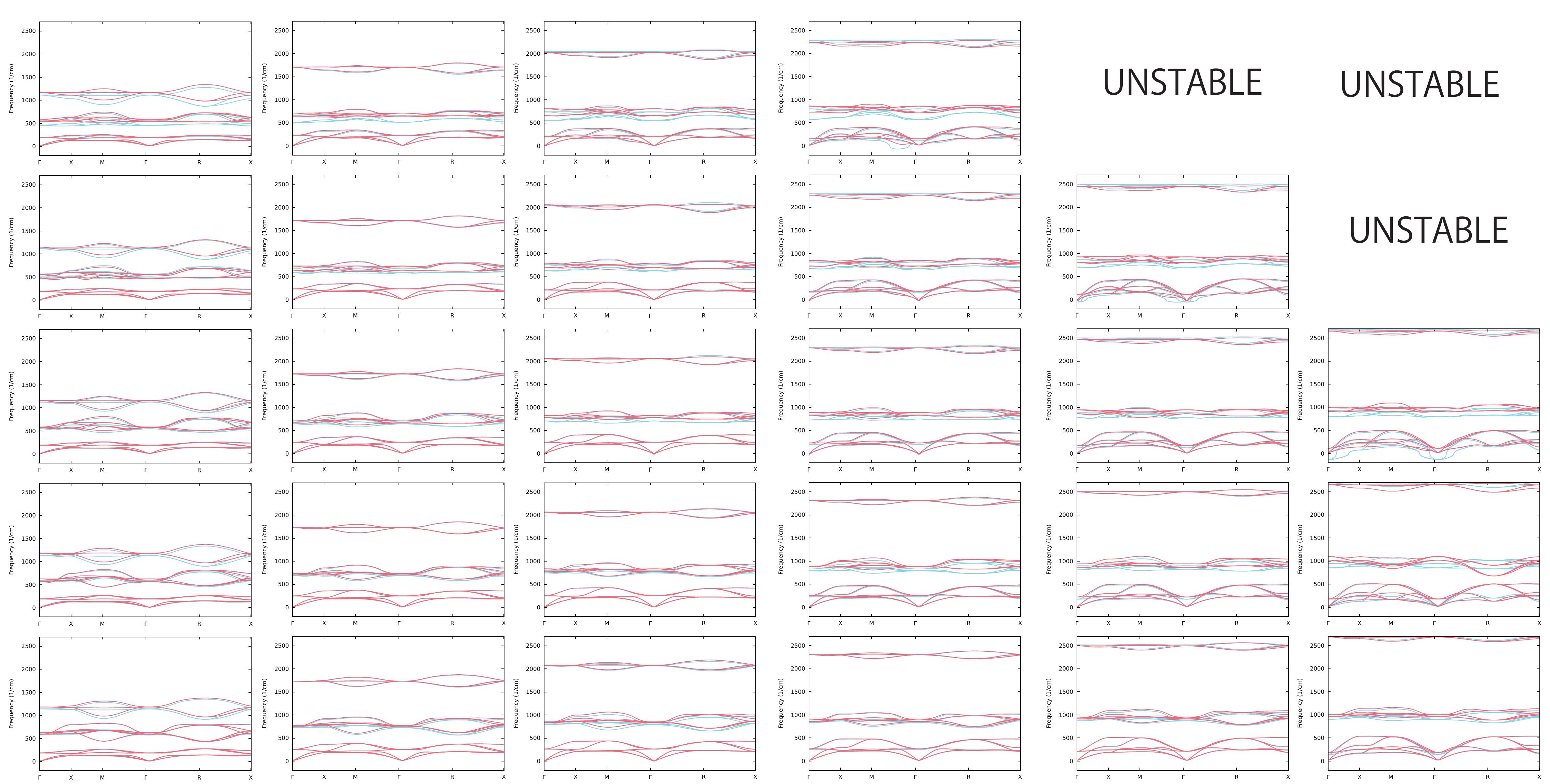"}
  \caption{Phonon dispersions of doped \ce{KMgH3}. The red blue lines are phonon frequencies 
  calculated within harmonic approximation. The blue lines are anharmonic phonon frequencies 
  obtained from self-consistent phonon (SCPH) calculations. The ordering is the same as in Fig. \ref{a2F}.}
\end{figure}

\begin{figure}[!ht]
  \subfloat[\label{nef}]{\includegraphics[width=0.5\columnwidth]{"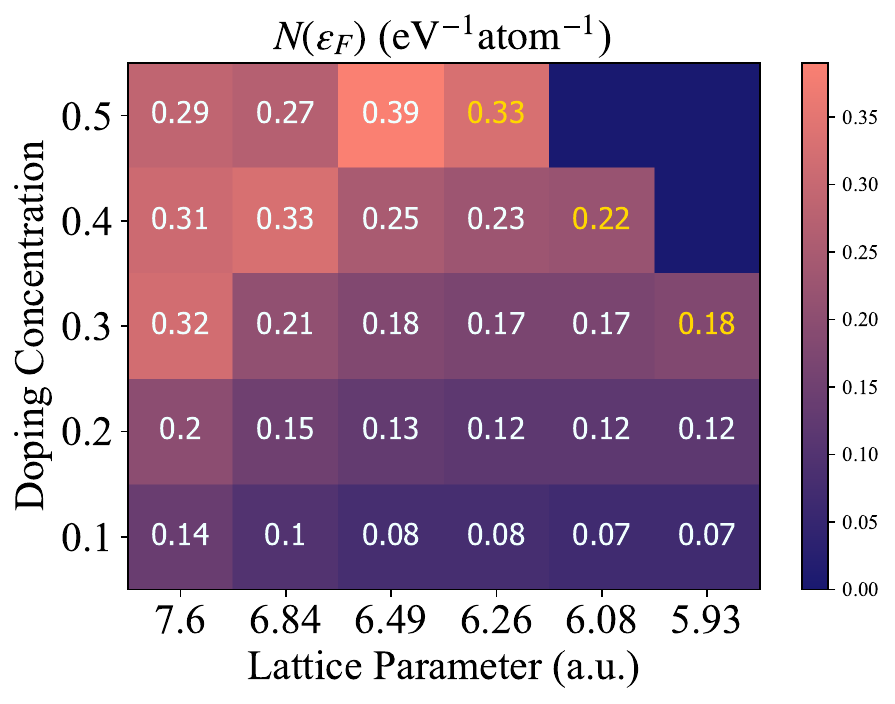"}}
  \subfloat[\label{wlog}]{\includegraphics[width=0.5\columnwidth]{"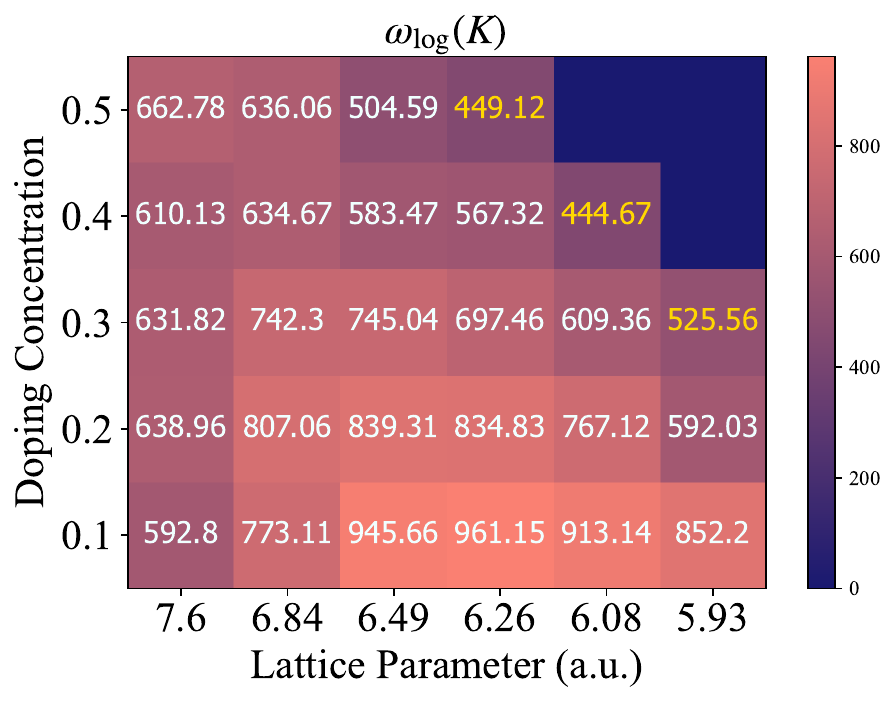"}}
  \caption{(a) Electronic density of states at the Fermi level $N(\epsilon_F)$. 
  (b) Logarithmic average of phonon frequencies $\omega_{log}$ of doped \ce{KMgH3}.}
\end{figure}
\newpage
\subsection{$\mathbf{K_2LiMgH_6}$}
\ce{K2LiMgH6} is a lithium-doped \ce{KMgH3}, which corresponds to a doping 
concentration of 0.5, where half of the Mg atoms are replaced by lithium atoms.
The crystal structure was optimized using the Vienna Ab Initio Simulation Package (VASP) 
\cite{KRESSE199615, PhysRevB.47.558, PhysRevB.59.1758, PhysRevB.54.11169}
with a $12\times12\times12$ $k$ grid under 35 GPa. Its transition temperature ($T_c$) 
was predicted, from SCDFT, to be 18.2 K (without spin fluctuation (SF)) and 0.7 K (with SF).
\begin{figure}[!ht]
  \includegraphics[width=0.8\columnwidth]{"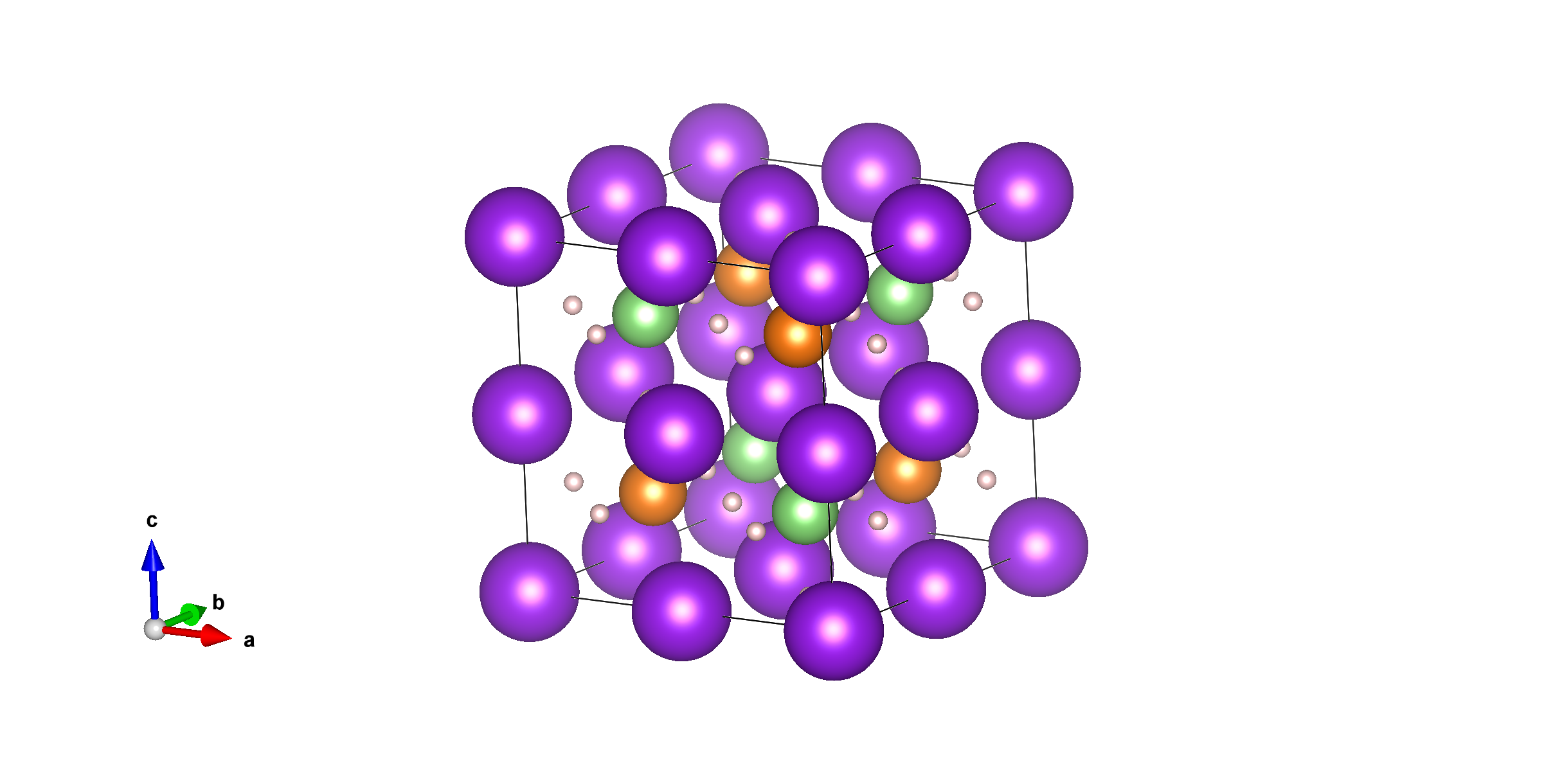"}
  \caption{Crystal structure of \ce{K2LiMgH6}. The purple, green, orange, and pink balls are 
  potassium, lithium, magnesium and hydrogen atoms, respectively.}
\end{figure}
\newpage
\begin{figure}[!ht]
  \centering
  \includegraphics[width=0.6\columnwidth]{"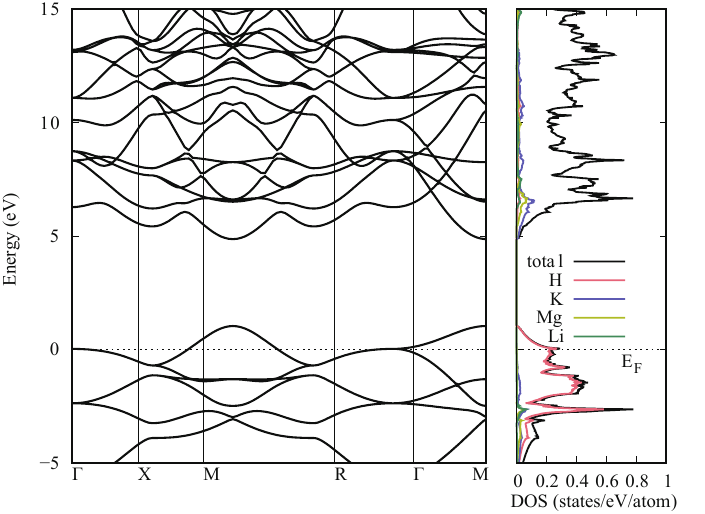"}
  \caption{Electronic bands structure of \ce{K2LiMgH6}.}
\end{figure}
\begin{table}[!ht]
  \centering
  \begin{tabular}{cccc}
    \hline
    \multicolumn{4}{c}{$Fm\bar{3}m$ at 35 GPa} \\
    \hline
    a, b, c [\AA] & 6.288552 & 6.288552 & 6.288552 \\
    $\alpha$, $\beta$, $\gamma$ & 90 & 90 & 90 \\
    H (24e) & 0.26059 & 0.50000 & 0.50000 \\
    Li (4b) & 0.50000 & 0.50000 & 0.50000 \\
    K (8c) & 0.25000 & 0.25000 & 0.25000 \\
    Mg (4a) & 0.50000 & 0.00000 & 0.50000 \\
    \hline
  \end{tabular}
  \caption{Calculated structure parameters for \ce{K2LiMgH6}.}
\end{table}

\subsection{Alkaline earth metal hydrides}
The alkaline earth metal hydrides \ce{XH3} do not possess dynamical stability at ambient pressure. 
Nevertheless, we can still estimate $\mu_s$ of them, since $\mu_s$ do not depend on phononic 
properties in the current scheme. They are all in the same $Fm\bar{3}m$ space group, and 
the lattice parameters are 4.71 \AA, 5.28 \AA, and 5.69 \AA, for X being Mg, Ca and Sr, respectively. 
The X atom is in 4a Wyckoff position and the coordinates are (0.00, 0.00, 0.00). 
The first H atom is in 4b Wyckoff position and the coordinates are (0.50, 0.00, 0.00), and 
another H is in 8c Wyckoff position with its coordinates being (0.25, 0.25, 0.75).
Calculated values of $\mu_s$ are listed in Table. \ref{mu_table}.

\begin{figure}[!ht]
  \centering
  \includegraphics[width=0.8\columnwidth]{"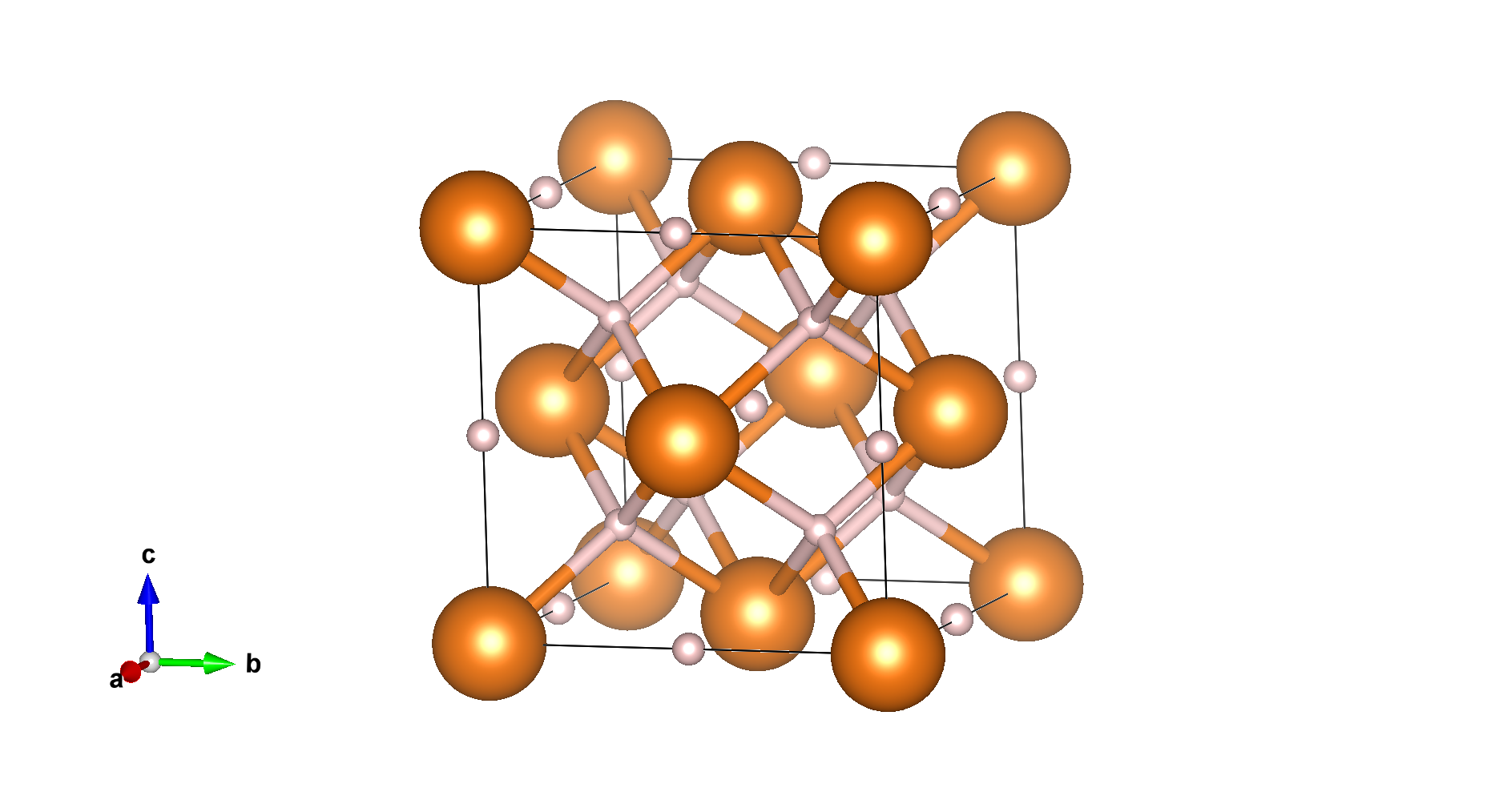"}
  \caption{The crystal structure of $Fm\bar{3}m$ \ce{XH3}. The orange balls are X atoms, 
  and the pink balls are H atoms.}
\end{figure}

\begin{table}[!ht]
  \centering
  \begin{tabular}{lcr}
    \hline
     & $N(\epsilon_F)$ [$\mathrm{eV}^{-1}\mathrm{atom}^{-1}$] & $\mu_s$ \\
    \hline
    \ce{MgH3} & 0.122 & 0.24 \\
    \ce{CaH3} & 0.169 & 0.30 \\
    \ce{SrH3} & 0.217 & 0.52 \\
    \ce{K2LiMgH6} & 0.237 & 0.67 \\
    \ce{KCdH3} & 0.074 & 0.034 \\
    \ce{KCdH3} (doped) & 0.144 & 0.15 \\
    \hline
  \end{tabular}
  \caption{Calculated $N(\epsilon_F)$ and the strength of 
  $\mu_s$. The doped \ce{KCdH3} is doped with 0.5 holes per unit cell.}
  \label{mu_table}
\end{table}

\begin{figure}[!ht]
  \centering
  \subfloat[\label{MgH3}]{\includegraphics[width=0.3\columnwidth]{"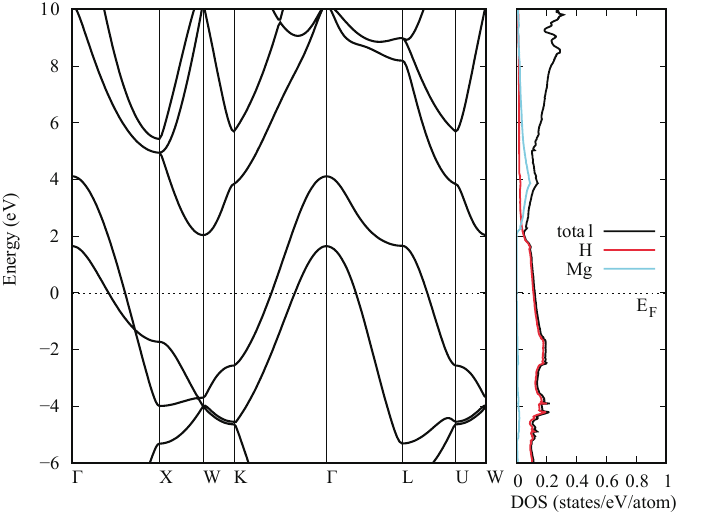"}}
  \subfloat[\label{CaH3}]{\includegraphics[width=0.3\columnwidth]{"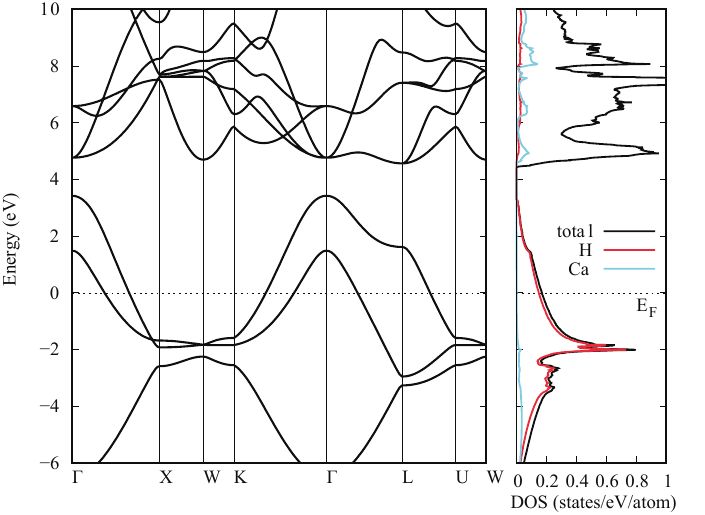"}}
  \subfloat[\label{SrH3}]{\includegraphics[width=0.3\columnwidth]{"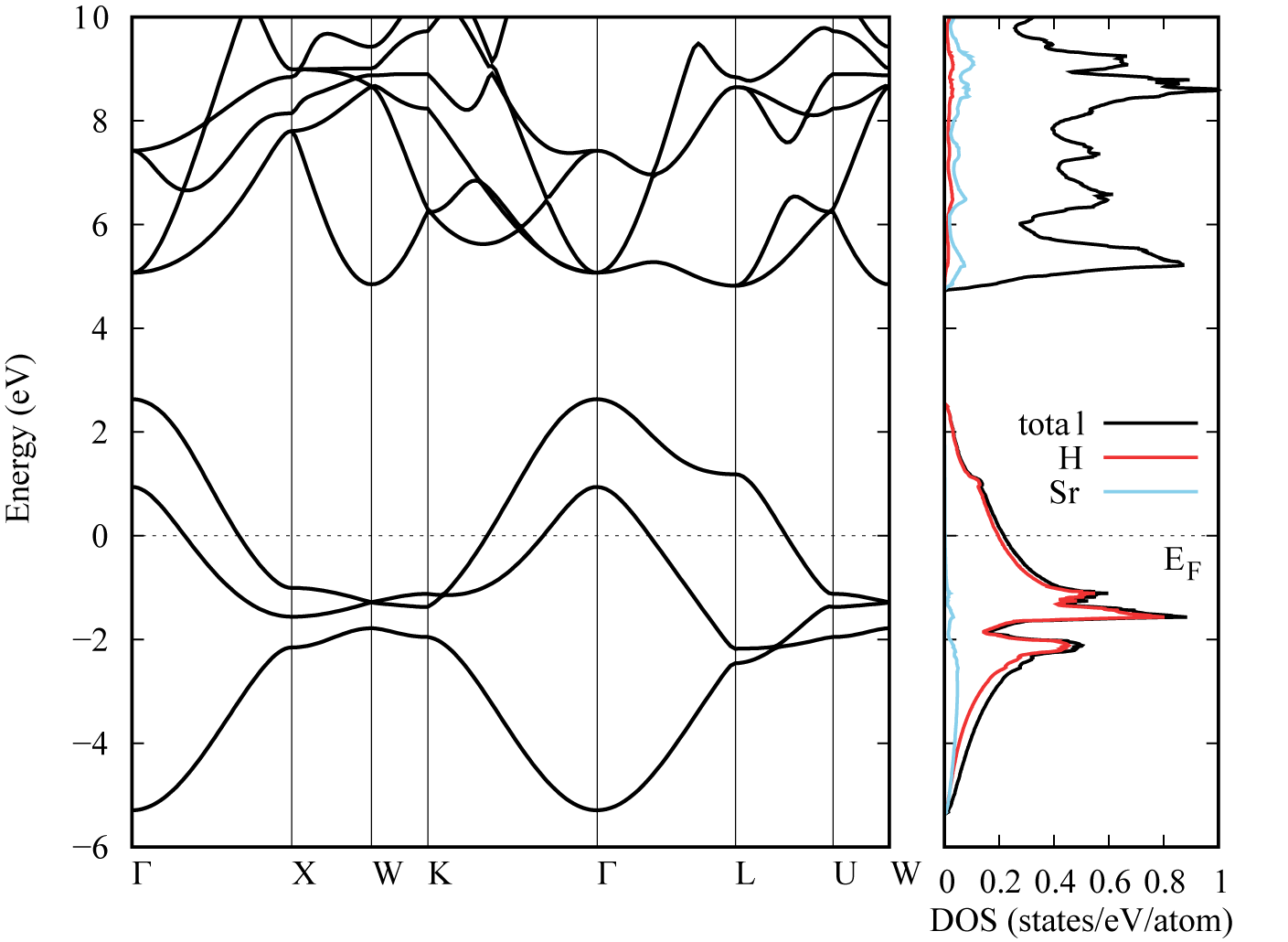"}}
  \caption{Electronic band structure of (a)\ce{MgH3}. (b)\ce{CaH3}. (c)\ce{SrH3}.}
\end{figure}

\subsection{Extra computational details}
The enthalpy calculations were performed using VASP, with a $12 \times 12 \times 12$ \textbf{k} point mesh. 
The energy bands that are used in the plots of 
electronic band structures were calculated with a $12 \times 12 \times 12$ \textbf{k} grid
with the Marzari-Vanderbilt-DeVita-Payne smearing method, and the broadening was set to be 0.005 Ry. 
Electronic energies used in calculations for DOS and in SCDFT for $\mu_s$ were calculated on a $16 \times 16 \times 16$ 
\textbf{k} grid in a non-self-consistent way using the charge densities from self-consistent calculations 
with $12 \times 12 \times 12$ \textbf{k} grids, and the optimized tetrahedron method were adopted in these 
calculations. Norm-conserved pseudopotentials \cite{VANSETTEN201839, PhysRevB.88.085117} with PBE exchange-correlation functionals were used. 
The exchange integrals of SF interactions were evaluated on a $4 \times 4 \times 4$ mesh. 
Crystal structures shown in this article are produced by VESTA\cite{Momma:db5098}.

\end{document}